\newif\ifdoubleblind\doubleblindfalse

\ifdoubleblind
\documentclass[3p,twocolumn,doubleblind]{elsarticle}
\else
\documentclass[3p,twocolumn]{elsarticle}
\fi
\usepackage{amsmath}
\usepackage{amssymb}
\usepackage{listings}
\usepackage[most]{tcolorbox}
\ifdefined\directlua
\usepackage{fontspec}
\newfontfamily{\timesnewroman}{Times New Roman}
\else
\usepackage[T1,T2A]{fontenc}
\let\timesnewroman\relax
\DeclareSymbolFont{T2Aletters}{T2A}{cmr}{m}{it}
\fi
\usepackage{url}

\usepackage{pgfplotstable}
\usepackage{booktabs}
\usepackage{siunitx}
\usepackage{colortbl}
\usepackage{hyperref}
\hypersetup{hidelinks}
\title{A new idea for RSA backdoors}
\author{Marco Cesati\fnref{orcid}}
\ead{cesati@uniroma2.it}
\affiliation{organization={DICII, University of Rome Tor Vergata},
                addressline={Via del Politecnico 1},
                postcode={00133},
                city={Rome},
                country={Italy}}
\fntext[orcid]{ORCID iD: \url{https://orcid.org/0000-0002-7492-3129}}
\def\theaffn{\relax} % Remove useless marker at left of address (single author)
\journal{arXiv}

\newif\ifrunexample\runexamplefalse
\ifrunexample\newtcolorbox{runexample}{sharp corners, leftright skip=.05\linewidth}\fi

\newcommand{\SSB}{SSB}
\newcommand{\TSB}{TSB}

\newcommand{\bitsize}[1]{\ensuremath{\ell({#1})}}

\newcommand{\abs}[1]{\ensuremath{\left|{#1}\right|}}
\newcommand{\set}[1]{\ensuremath{\left\{{#1}\right\}}}

\newcommand{\cmodl}[3]{\ensuremath{%
    {#1}\nolinebreak\equiv\nolinebreak{#2}\pmod{#3}}}
\newcommand\modequiv[1]{\ensuremath{%
        \equiv_{\raisebox{-2pt}{\scriptsize ${#1}$}}}}
\newcommand{\cmods}[3]{\ensuremath{%
        {#1}\nolinebreak\modequiv{#3}\nolinebreak{#2}}}
\let\omod=\bmod

\let\cmod=\cmods
\let\ncmod=\ncmods
\newcommand{\MSB}[2]{\ensuremath{\left.{#1}\right\rceil^{#2}}}
\newcommand{\LSB}[2]{\ensuremath{\left.{#1}\right\rfloor_{#2}}}

\newcommand*\topfigrule{\hrule\kern-0.4pt\relax}
\newcommand*\botfigrule{\hrule\kern-0.4pt\relax}
\begin{document}
\begin{abstract}
        This article proposes a new method to inject backdoors in RSA and other
        cryptographic primitives based on the Integer Factorization problem for
        balanced semi-primes.  The method relies on mathematical congruences
        among the factors of the semi-primes modulo a large prime number, which
        acts as a ``designer key'' or ``escrow key''.  In particular, two different
        backdoors are proposed, one targeting a single semi-prime and the
        other one a pair of semi-primes. The article also describes the results
        of tests performed on a SageMath implementation of the backdoors.
\end{abstract}
\begin{keyword}
        RSA\sep Backdoor\sep Escrow key\sep Implicit Factorization Problem\sep Integer Factorization
\end{keyword}

\maketitle
\section{Introduction}
\label{s:intro}

Impairing the robustness of cryptographic applications is a sensitive topic.
The interest on direct attacks, vulnerabilities, and backdoors for all
currently used ciphers is certainly justified by economic or geopolitical
reasons. If a vulnerable implementation of a cryptographic algorithm is
surreptitiously distributed, an ``evil'' actor or a national security agency
might get easy access to any sort of sensitive and precious information. On the
other hand, there could be ``legal'' actors that openly mandate or encourage the
adoption of cryptographic implementations that include backdoors in order to
realize ``key escrow'' mechanisms. For instance, a national country might
legislate that judiciary representatives should always be able to recover any
kind of encrypted communication involved in a criminal case.

Up to a few years ago, it was only conjectured \cite{filiol2017} that major
security agencies were able to decrypt a large portion of the world's encrypted
traffic, mainly thanks to vulnerabilities hidden in pseudo-random generators or
major cryptographic algorithms and applications. Some examples of
this practice might be the Hans B\"uhler case in 1994 \cite{strehle1994}, the
Dual-EC algorithm proposed in 2004 by US National
Institute of Standards and Technologies \cite{NIST2006,bernstein2016}, and
perhaps the OpenBSD backdoor incident emerged in 2010
\cite{leyden2010,paul2010}. However, in the last few years many government
bodies openly talk about enforcing by law ``responsible encryption'' or
``exceptional access to encrypted documents'' \cite{rosenstein2017,levy2018},
which are essentially more palatable words for ``escrow key'' and
``backdoors''. 

Moreover, even the approach to backdoor construction is changed in the last
years.  While in the past the focus was mainly on weaknesses in pseudo-random
generators or software implementations that might allow an attacker to predict
some secret data of the target users, nowadays the emphasis is on theoretical
backdoors based on mathematical properties of the cryptographic primitives.
Perhaps, the main reason for this new approach is that it is very difficult to
discover a mathematical backdoor by just looking at the cryptographic
algorithm: for example, Bannier and Filiol \cite{bannier2017} showed in 2017
how a block cipher similar to AES can be devised so that it includes, by
design, a hidden mathematical backdoor that allows a knowledged attacker to
effectively break the cipher and recover the key.

Evil actors and legal actors pursue very different goals, which justify the
adoption of very different backdoor mechanisms. An evil actor is primarily
concerned with how convenient is triggering the backdoor, and secondarily
with how well the backdoor mechanism is hidden to the final user; if
that mechanism also impairs the security of the cipher is not crucially
important. Thus, a backdoor introduced by an evil actor might even be a
vulnerability hidden in a cipher implementation such that anyone knowing about
its existence could easily break the cipher and recover the encrypted messages.
For instance, a mechanism that can be easily exploited might be based on a
semi-prime generator that select just one of the primes at random, while the
other prime is fixed.  The Euclidean algorithm applied to two different
vulnerable semi-primes outputs the fixed prime, thus anyone can easily break
the cipher even if the fixed prime is not known in advance. Perhaps not
surprisingly, there are in Internet a lot of very weak public keys
\cite{heninger2012, bernstein2013}.
On the other hand, a legal actor does not want to significantly impair the
security of a cryptographic algorithm, because the final users might just
refuse to adopt an insecure cipher. A backdoor introduced by a legal actors is
likely a vulnerability embedded in a cryptographic implementation that allows
only ``authorized'' actors to decipher the encrypted messages without knowing
the private keys of the final users.  Usually, this means that the retrieval of
the encrypted messages can be done only if the actor knows a secret escrow
key related to the backdoor itself.

Among the most widespread cryptographic algorithms, RSA \cite{rivest1978}
likely deserves special consideration, because it is conveniently used to
protect any kind of sensitive data transmitted over the Internet.  While it is
commonly believed that RSA has been properly designed and that, by itself, it
does not contain hidden vulnerabilities, a large number of attacks to RSA have
been proposed since its invention.  These attacks span from directly factoring
the semi-prime in the public key to exploiting weaknesses in the generation
algorithm for the prime factors; for a survey, see \cite{mumtaz2019}.
Furthermore, several RSA's backdoors have been proposed: they are specially
crafted values in RSA parameters that allow a knowledged attacker to recover
the private key from publicly available information.  For a in-depth discussion
of several RSA's backdoors see \cite{arboit2008}. 

In this work we propose a new idea to inject backdoors in RSA key generators,
which was loosely inspired by the concept of ``implicit hints'' of May and
Ritzenhofen~\cite{may2009} in pairs of semi-primes. On the other hand, our idea
differs significantly from the backdoors based on implicit hints and, as far as
we know, from any other published backdoor proposal.

More specifically, May and Ritzenhofen proposed the Implicit Factorization
Problem (IFP), which is based on the premise that two or more semi-primes
having factors sharing some common bits can be factored with some variants of
the Coppersmith's algorithm~\cite{coppersmith1996,coppersmith1996b}.  The
authors stated that ``[\ldots] one application of our result is malicious key
generation of RSA moduli, i.e. the construction of backdoored RSA moduli''.  In
our opinion, however, a backdoor based on shared bits, as described in
\cite{may2009}, is not really effective for RSA. In fact, it is practically
not possible to exploit this backdoor in large ``balanced'' semi-primes, such as
those used in currently used RSA moduli, because the time required by the
Coppersmith's algorithm to factor a semi-prime grows exponentially when the
difference in the size of the factors becomes smaller. Moreover, this
vulnerability is self-evident to anyone looking at the factors, because there
would be a long run of identical bits in the two values, which means that the
backdoor cannot be easily concealed to the owner of the private keys.

Our new idea is the following: rather than prescribing that the bit-expansions
of the factors include a long run of \emph{identical} bits, we impose that the
bit-expansions include portions of \emph{correlated} bits, where the
correlation is bound to a secret designer key not known to the owner of the keys.
In practice, we impose some mathematical conditions on the values of the
factors as congruences modulo a large prime (of nearly the same size of the
factors), which acts as the designer key.

Following the IFP approach in \cite{may2009}, we firstly devised a backdoor
(named \TSB) based on mutual correlations between the factors of two distinct
semi-primes. Afterwards we devised a simpler backdoor (named \SSB) based on the
same idea but suitable for injecting a backdoor in a single semi-prime. 
The backdoors can be applied to RSA or to any other cipher whose security is
based on the difficulty of the integer factorization of the semi-primes.

A key difference with the IFP approach is that in order to trigger the
backdoors, that is, in order to factor the semi-prime(s) by exploiting the
designer key, there is no need to apply some variant of the Coppersmith's
algorithm. Therefore, if the value of the designer key is known, factoring the
semi-prime(s) is easy and efficient.  On the other hand, if the designer key is
not known, there seems to be no efficient way to factor the semi-prime(s).
Moreover, without the designer key, there seems to be no efficient way to detect
the \emph{existence} of the backdoor, even when looking at the distinct prime
factors of the semi-prime(s).

The rest of the article is organized as follows. In
Section~\ref{s:preliminaries} we define some mathematical notation and
introduce the basic RSA algorithm. In Section~\ref{s:relwork} we present the prior
works related to RSA backdoors and the Implicit Factorization Problem (IFP). 
In Sections~\ref{s:ssb} we discuss our simpler backdoor, \SSB, while in
Section~\ref{s:tsb} we discuss the more sophisticated backdoor for a pair of
semi-primes, \TSB. Finally, in Section~\ref{s:conc} we draw some conclusions
from this work.

\section{Preliminaries}
\label{s:preliminaries}

Let us establish some notations: $\cmodl{a}{b}{c}$ denotes the relation in
which $a-b$ is a multiple of $c$; often we will use the shorter notation
$\cmods{a}{b}{c}$.  The notation $a\omod b$ denotes the operation remainder of
the division $a/b$; hence, $\cmod{a}{(a\omod c)}{c}$ and $0\leq a\omod c<c$.

If $N\geq 0$ is an integer, its size in bits is defined as
$\bitsize{N}=\max\set{1,\left\lceil\log_2(N+1)\right\rceil}$.  We write
$x\simeq y$ if $x$ and $y$ are equal or differ by at most one, while we write
$x\approx y$ if $x$ and $y$ differs by a value negligible with respect to the
sizes of $x$ and $y$. If $N\approx 2^n$, with $n$ large, then
$\bitsize{N}\simeq\log_2 N$, that is, we may consider both $\bitsize{N}$ and
$\log_2 N$ to be approximately equal to $n$, ignoring a $\pm 1$ difference in
size.

If $h$ is an integer, \MSB{h}{k} denotes the $k$ most significant bits of $h$
(a value from 0 to $2^k-1$), while \LSB{h}{k} denotes the $k$ less significant
bits.

A \emph{semi-prime} is a number $N$ such that $N=p\,q$ where $p$ and $q$ are
primes. Therefore, $\bitsize{N}\simeq\bitsize{p}+\bitsize{q}$.  If
$\bitsize{p}\simeq\bitsize{q}$, then the semi-prime is said to be
\emph{balanced}.  In the following we consider also sequences of
semi-primes $N_i=p_i\,q_i$ ($i=1,2, \ldots$) having common size
$n=\bitsize{N_i}$, for every~$i$; furthermore, the primes $q_i$ have common
size $\bitsize{q_i} =\alpha$; it follows that all primes $p_i$ have the same
size $n-\alpha$.

The RSA public key cryptosystem has been invented by Rivest, Shamir, and
Adleman \cite{rivest1978} in 1977. In its simplest form, the algorithm is based
on a balanced semi-prime $N=p\,q$ and a couple of exponents $e$, $d$ such that
$\gcd(e,\phi(N))=1$ and $\cmodl{ed}{1}{\phi(N)}$.  Here $\phi(N)$ denotes the
Euler's totient function, which can be easily computed as $(p-1)(q-1)$ if the
prime factors $p$ and $q$ are known. Theoretically, the value of $e$ could be
random, while the value of $d$ can be computed from $e$ and $\phi(N)$ by using
the Extended Euclidean algorithm. The pair $(N,e)$ is the ``public key'' of
RSA, and the encryption function is $M^e\omod N$. Either the pair $(p,q)$ or
the pair $(N,d)$ is the ``private key'', and the decryption function is
$(M^e)^d\omod N \equiv M^{s\phi(N)+1}\equiv M(M^{s\phi(N)}) \equiv
\cmodl{M(1^s)}{M}{N}$. Of course, factoring $N$ allows an attacker to recover
the private key from the public key, because from $p$ and $q$ we can compute
$\phi(N)$ and then $\cmodl{d}{e^{-1}}{\phi(N)}$.

\section{Related work}
\label{s:relwork}

Many authors proposed to classify backdoors embedded in cryptographic
applications according to several, different criteria. Following
\cite{markelova2021}, we consider three types of backdoors: (1) weak backdoors,
(2) information transfer via subliminal channels, and (3) SETUP mechanisms.
Weak backdoors are based on modifications of the cryptographic protocol such
that it would be possible to anyone to break the cipher and recover the secret
data. Vulnerabilities falling under the information transfer via subliminal
channels category allow an attacker to exploit the cryptographic protocol in
such a way to create a hidden communication channel that cannot be intercepted
or unambiguously detected.  Finally, SETUP (Secretly Embedded Trapdoor with
Universal Protection) mechanisms create vulnerabilities in the cryptographic
protocols that cannot easily exploited by third-party attackers.

SETUP mechanisms have been firstly proposed by Young and Yung
\cite{young1996,young1997} in 1996: they coined the term ``kleptography'' to
denote the usage of cryptographic primitives in order to design ``safe''
backdoors in other cryptographic protocols. Following the classical distinction
between asymmetric and symmetric cryptography, SETUP mechanisms can lead to
\emph{asymmetric backdoors} and \emph{symmetric backdoors}.

In an {asymmetric backdoor} the information required to recover the
encrypted messages is protected by an asymmetric cipher.  Usually, this means
that some data that allows an actor to recover any user private key are
encrypted with the public key of the designer of the RSA implementation and
stored inside the corresponding user public key. Any actor that knows the
corresponding designer private key may extract the data from the user public
key and decipher them to recover the user private key.  Observe that in this
case the RSA implementation is ``tamper resistant'': even reverse engineering
cannot reveal the designer private key.

In a {symmetric backdoor}, on the other hand, the designer key that allows
an actor to recover the user private key from the user public key is stored in
some form inside the RSA implementation itself. To be secure and undetected,
the RSA implementation (perhaps, a physical device) must be ``tamper proof''.

Existing RSA backdoors may also be categorized according to the place where the
backdoor's specific data are stored: either in the semi-prime $N$ alone, or
also in the exponent $e$ of any public key $(N,e)$. ``Exponent-based''
backdoors are somewhat easier to devise, because $e$ could theoretically be any
random value coprime with $\phi(N)$. However, most RSA implementations make use
of special fixed values for the public exponent, such as small values or values
having small Hamming weight, in order to improve the efficiency of the RSA
algorithm.  Thus, exponent-based backdoors cannot be easily hidden to the final
user, and can be perceptively slower than honest RSA implementations. Backdoors
embedded in the public key's semi-prime does not limit the choice of the public
exponent, however they must address a crucial problem: how to encode
information about the factorization of the semi-prime in the semi-prime itself,
in such a way that the information is encrypted with a secret key and,
possibly, the pair $(p,q)$ is indistinguishable by a pair of primes generated
by a honest RSA implementation. 

In this work we propose two backdoors embedded in the semi-primes of the RSA's
public keys; as a matter of fact, the backdoors apply to any cryptographic
protocol based on the integer factorization of semi-primes. Therefore, we don't
discuss at length related work concerning exponent-based backdoors; examples
can be found, for example, in \cite{howgrave2001,crepeau2003,arboit2008,
sun2009}

\subsection{Symmetric backdoors}

The proposed \SSB\ algorithm implements a symmetric backdoor, because the
escrow key is fixed and hard-cabled in the hardware or software device 
that generates the vulnerable semi-primes. As we shall see, \TSB\ might
be considered both a symmetric or an asymmetric backdoor.

The first RSA backdoor has been proposed by Anderson \cite{anderson1993} in
1993.  It is a symmetric backdoor embedded in the public key's semi-prime:
let $\beta$ be a $m$-bit secret prime (the ``backdoor key''), and let
$\pi_\beta$ and $\pi'_\beta$ be pseudo-random functions that, given a seed in
argument, produce a $(n-m)$-bit value (in the original article, $n=256$ and
$m=200$).  For any vulnerable $2n$-bit semi-prime $N=pq$, let
$t,t'<\sqrt{\beta}$ be $(m/2)$-bit random numbers coprime with $\beta$, and let
$p=\pi_\beta(t)\cdot \beta+t$ and $q=\pi'_\beta(t')\cdot\beta+t'$.  Given $N$
and $\beta$, it is possible to compute $tt'=N\omod\beta $, then factor the
$m$-bit number $tt'$, and finally compute $p$ and $q$.  Kaliski
\cite{kaliski1993} proves that it is possible to discover the backdoor by
either computing the continued fraction $p/q$, because the expansion likely
contains an approximation of the fraction $\pi_\beta(t)/\pi'_\beta(t')$, or by
finding a reduced basis of a suitable lattice built on the primes of two
vulnerable moduli. He also shows that the backdoor can be detected by the
lattice method when 14 or more non-factored vulnerable moduli are available.  It
is easy to observe that Kaliski's detection algorithm can be easily defeated by
introducing a ``dynamic backdoor key'' whose exact value depends, for instance,
on an incremental counter.  However, another drawback of Anderson's backdoor is
that $m\approx 3/4\cdot n$, hence triggering the backdoor for currently used
public key sizes might require factoring a too large integer.

Our first proposed backdoor, \SSB, is similar to Anderson's construction, in
that triggering the backdoor involves as first step computing the remainder of
the integer division of the semi-prime and the designer (escrow) key. However,
a key difference with Anderson's idea is the form of the primes $p$ and $q$,
which allows \SSB\ to escape detection by Kaliski's algorithms and to avoid
factoring a large integer when exploiting the backdoor.

In 2003, Crep\'{e}au and Slakmon \cite{crepeau2003} presented, among several
others exponent-based backdoors, a semi-prime-based backdoor that relies on
Coppersmith's attack \cite{coppersmith1996b} and encrypts the factor~$p$ in the
RSA modulus $N=pq$ in such a way that the bits in $\MSB{N}{n/8}$ have the
correct distribution for a random semi-prime, while the middle $n/4$ bits
of~$N$ are an encryption, via a pseudo-random function~$\pi_\beta$, of
$\MSB{p}{n/4}$. Our proposed backdoors use a entirely different mechanism and
do not rely on Coppersmith's attack, which means that they can be efficiently
exploited even on very large balanced semi-primes.

In 2008, Joye \cite{joye2008} studied the performances of generating a
semi-prime $N$ in which some bits are prescribed; he developed as an example a
RSA symmetric backdoor based on the Coppersmith's attack in which some of the
bits of~$p$ are encrypted in~$q$. While this study is relevant when analyzing
the generation times of any semi-prime backdoor, their proposal is entirely
different than the present one.

\iffalse
In 2009, Sun, Wu, and Yang \cite{sun2009} proposed a symmetric backdoor
embedded in the semi-prime $N=pq$ exposing a vulnerable exponent of the RSA
modulus. Actually, the authors make use of an optimized RSA method based on the
Chinese Remainder Theorem \cite{quisquater1982} that split the exponent in two smaller
values. The recovery procedure extracts the vulnerable exponents from $N$ by
using a lattice-based method.
Another interesting backdoor proposed in \cite{sun2009} is the following: when
generating $N=pq$, the prime $q$ is forged so that $q-1$ satisfies $\beta
d-K( p-1)(q-1)=1$, where $\beta$, the designer secret key, is a prime
smaller than $q-1$ and larger than $(q-1)/2$, and $K$ is a random value
such that $K(p-1)$ is smaller than $d$ and it is larger than $d/2$.  The
actor tries all possible values of $K$ to compute
$\cmod{p+q}{N+1+K^{-1}}{\beta}+j\beta$, for $j=0,1,2$, and
recognizes the right value for $p+q$ by checking that
$\cmod{\alpha^{N+1-(p+q)}}{1}{N}$ for random $\alpha$. Once $p+q$ is known,
it is possible to factor $N$ by computing $p-q=\sqrt{(p+q)^2-4N}$.
The authors suggested $\bitsize{K}=40$ bits.
\fi

The symmetric backdoor proposed by Patsakis \cite{patsakis2012} in 2012 is
based on yet another idea: the parameterized, randomized backdoor
algorithm decomposes an integer as sum of squares in a way depending on a
designer's secret parameter. The backdoor consists in imposing that the
semi-prime, once decomposed by using the secret parameter, can be easily
solved by a nonlinear system whose solutions are properly bounded.

In 2017, Nemec, Sys, and others \cite{nemec2017} exposed a critical
vulnerability (perhaps unintentional) in the key generation algorithm of the
\emph{RSALib} library, which is written, adopted, and distributed to third
parties by Infineon, one of the top producers of cryptographic hardware
devices. This work raised much interest because the flaw was already present in
devices produced in 2012 and the total number of affected devices, and
consequently vulnerable keys, is huge.
In any $N=pq$ generated by the flawed \emph{RSALib}, all primes $p$ and $q$ have
the form $k\cdot M_t+(65537^a\omod M_t)$, where $M_t$ is the \emph{primorial}
number composed by the product of the first $t$ primes, and $k$, $a$ are random
integers. The values of $t$ for semi-primes of bit length $n=512$, $1024$, $2048$, and
$4096$ are, respectively, $t=39$, $71$, $126$, and $225$. This means that the
number of truly random bits in each of the primes is reduced, respectively, to
98, 171, 308, and 519.
In order to find the factors of a vulnerable semi-prime, a variant of the
Coppersmith's attack is used: it is possible to efficiently factor $N=pq$
when the value $p\omod M$ is known.  Hence, the recovering procedure determines
a suitable divisor $M$ of $M_t$ of size $\bitsize{M}\geq n/4$ (to reduce the
search space for $a$), guesses an exponent $a$, computes $67537^a\omod M$, and
factors $N$. 
It is also easy to verify whether a given key is flawed: $N$ is likely
vulnerable if the discrete logarithm $\log_{65537} N\omod M_t$ exists.
Actually, this logarithm can be easily computed by the Pohlig-Hellman algorithm
\cite{pohlig1978} because $M_t$ is the product of many small consecutive
primes. Hence, ROCA belongs arguably to the weak backdoor category.

\subsection{Asymmetric backdoors}

The proposed \TSB\ algorithm can be used to implement both symmetric and
asymmetric backdoors. In fact, \TSB\ makes use of an embedded designer key, but
also generates two distinct semi-primes. If both semi-primes are used to build
two distinct public keys, both available to a third-party attacker, then
tampering with the \TSB\ device may expose the designer key and break the keys.
On the other hand, \TSB\ can be used to generate a public key (from one of the
generated semi-primes) and a dedicated escrow key composed by the hard-coded
large prime inside the device and the second semi-prime, which must be considered
as the designer's secret key. This is a reasonable scenario for cryptographic
keys used in a highly-secure work environment. In this second case, \TSB\ must
be considered an asymmetric backdoor, because tampering with the device is not
enough to break any key already generated.

The first examples of asymmetric backdoors proposed by Young and Yung
\cite{young1996} in 1996 were exponent-based. However, that article also
includes the description of a asymmetric
\iffalse
In 1996, Young and Yung \cite{young1996} introduced the idea that a
cryptographic protocol may be used to inject a vulnerability in another
cryptographic protocol.  In particular, a \emph{kleptographic} backdoor is a
mechanism implemented by means of an asymmetric cryptographic protocol aimed to
subvert the security of another cryptographic protocol.  As a first example,
the authors showed an asymmetric, exponent-based RSA backdoor implemented by
means of RSA.  The backdoor designer defines its own RSA public key
$(N'=p'q',e')$ and private key $(p',q',d')$, where $\bitsize{N'}=n/2$. The
semi-prime $N=pq$ is generated as usual (with $\bitsize{p}=\bitsize{q}=n/2$),
however the exponent $e$ is not chosen at random, but rather it is set to
$e=p^{e'}\omod N'$. If provided with the designer's private key, the actor
can simply extract the factor $p$ from the exponent $e$. Observe that
$\bitsize{e}=n/2$, hence the security of the RSA algorithm is severely
impaired by this backdoor.  In \cite{young1996} is also presented an asymmetric,
\fi
semi-prime-based backdoor named PAP, for ``Pretty Awful Privacy''.  The
backdoor designer defines a designer's RSA public key $(N'=p'q',e')$ and
private key $(p',q',d')$, where $\bitsize{N'}=n/2$.
Let $F_K$ and $G_K$ be invertible functions depending on a fixed key $K$ that
transform a seed of $n/2$ bits in a pseudo-random value of $n/2$ bits. In order
to create a backdoor in a RSA moduli, the designer first chooses a prime $p$ of
bit length $n/2$ at random, then searches the smallest value $K$ such that
$\rho=F_K(p)<N'$. $\rho$ is then encrypted as $\rho_2=G_K(\rho^{e'}\omod N')$.
The RSA semi-prime $N$ results from the search of a prime $q$ such that the
$n/2$ most significant bits of $N=pq$ coincide with $\rho_2$. The attacker can
easily break the public key by extracting $\rho_2$ from $N$, then starting an
exhaustive search of the value for $K$ that, when applied to the inverse
permutations $G_K^{-1}$ and $F_K^{-1}$, permits to extract the proper factor
$p$ using the RSA private key $(p',q',d')$.

In a series of articles published between 1997 and 2008, Young and Yung
\cite{young1997, young2005a,young2005b,young2008} proposed several
kleptographic backdoors for RSA using different cryptographic algorithms 
for embedding the factor $p$ in $N$.
\iffalse
Specifically, in \cite{young1997} the backdoor PAP2 is embedded in the RSA
semi-prime via the ElGamal protocol \cite{elgamal1985}, that is, encrypting $p$
in $N$ is based on a Diffie-Hellman key exchange.  In \cite{young2005a} the
backdoor PP, for ``Private Primes'', is based on Rabin's cryptosystem
\cite{rabin1979}; it also differs from the one described in \cite{young1997}
because it uses non-volatile memory to store the number of generated backdoored
keys so as lower the probability of producing the same key twice.  In
\cite{young2005b} the encryption of the factor $p$ inside the semi-prime $N$ is
achieved by means of elliptic curve Diffie-Hellman key exchange.  In 2008,
Young and Yung \cite{young2008} revisited the backdoor proposed in
\cite{young2005b} and implemented it on the OpenSSL library. After some
optimization effort, this implementation has been made faster than the original
OpenSSL RSA key generation methods.
\fi

In 2010, Patsakis \cite{patsakis2010,patsakis2012} proposed yet another kleptographic
mechanism that relies on Coppersmith's attack and forges $p$ and $q$ so that
the most significant bits of both of them are of the form $(a+r)^{e'}\omod N'$,
where $a$ is a secret design parameter, $r$ is a random value, and $(N',e')$ is
the designer's asymmetric public key.

In 2016, W\"uller, K\"uhnel, and Meyer \cite{wuller2016} proposed a RSA
backdoor called PHP, for ``Prime Hiding Prime'', in which the information
required to factor $N$ is hidden in $N$ itself. The idea is to select a
prime $p$ such that $q=(p^{e'}\cdot p^{-1})\omod N'$ is a prime, where
$(N',e')$ is the RSA public key of the designer. To factor $N=pq$, the
actor computes $\cmod{N^{d'}}{(p\cdot p^{e'}\cdot
p^{-1})^{d'}\modequiv{N'}p}{N'}$.  An improvement of PHP, called PHP', is also
described in \cite{wuller2016}: here $q=(s^{e'}\cdot p^{-1})\omod N'$, where
$s$ is the concatenation of $n/4$ random bits and $\LSB{p}{n/4}$.
Half of the bits of $p$ are enough to recover the factorization
of $N$ thanks to the Coppersmith's attack.

Markelova~\cite{markelova2021} revisited Anderson's idea for a symmetrical
backdoor and devised SETUP mechanisms that protect the backdoor by means of
some public-key algorithms, in particular based on discrete logarithm problems
on both finite fields and elliptic curves. The author also presented a SETUP
backdoor exploiting the Chinese Remainder theorem. The article~\cite{markelova2021}
also includes a discussion of the similarities of these SETUP backdoors with the
ROCA backdoor.

\subsection{The Implicit Factorization Problem}

In 1985, Rivest and Shamir \cite{rivest1985} introduced the \emph{oracle
complexity} as a new way to look at the complexity of the factorization problem
(and the related RSA attack): they showed that the semi-prime $N$ can be
factored in polynomial time if an oracle provides ${3}/{5}$ of the bits of
$p$. In 1996, Coppersmith \cite{coppersmith1996,coppersmith1996b} improved the
oracle complexity by showing that an explicit ``hint'' about the top half bits
of $p$ are sufficient for factoring $N$ in polynomial time.  In particular,
Coppersmith described some algorithms based on lattice reduction and the LLL
procedure \cite{lenstra1982} to find small integer roots of univariate modular
polynomials or bivariate integer polynomials. Later
\cite{howgrave1997,coron2004}, these algorithms have been reformulated in
simpler ways and heuristically extended to the multivariate polynomial case. 

The seminal article~\cite{may2009} focusing on ``implicit hints'' has been
published in 2009 and it is due to May and Ritzenhofen. An oracle gives an
implicit hint when it does not output the value of some bits of one of the
factors of the semi-prime; rather, the oracle outputs another semi-prime whose
primes share some bits with the primes of original semi-prime. The authors
formally introduced the Implicit Factorization Problem (IFP), and showed that
two semi-primes $N_1$ and $N_2$ can be factored in time $O(n^2)$ if
$\LSB{p_1}{t}=\LSB{p_2}{t}$, with $t\geq 2\alpha+3$. The
algorithm is based on a lattice reduction: the search for the unknown primes
$q_i$ is reduced to a search for a basis of a suitable lattice by means of the
quadratic Gaussian reduction algorithm. This result implies that only highly
imbalanced semi-primes can be factored, because
$\bitsize{q_1}=\bitsize{q_2}=\alpha$, hence $\bitsize{p_i}>2\,\bitsize{q_i}$.
The authors also extended this result to $k>2$ semi-primes, and showed that a
polynomial algorithm based on the LLL algorithm \cite{lenstra1982} exists if
$t\geq \alpha k/(k-1)$. For the balanced case this result is not useful,
because it means that all $p_i$ primes are identical, hence they can be easily
recovered by the Euclidean algorithm.  However, the authors also showed that
their method can be used to factor $k$ balanced semi-primes when some
additional conditions are satisfied and $n/4$ bits are discovered by brute
force.

In the following years many articles improved and extended the results of May
and Ritzenhofen \cite{sarkar2009,faugere2010,sarkar2011,kurosawa2013,lu2013,
sakai2014,peng2014,nitaj2015,nuida2015,lu2016,peng2015,harasawa2016,wang2018,
zhukov2018,sun2019,zheng2019,zheng2019b}. See also a survey \cite{lu2018}
published in 2018.

All attacks and vulnerabilities based on these results assume that the factors
of vulnerable semi-primes share some identical bits.  From a practical point of
view, backdoors relying on shared identical bits cannot be easily concealed to
anyone looking at the factors, that is, the private key. Furthermore, all the results
cited in this section are based on some variants of Coppersmith's algorithms
\cite{coppersmith1996, coppersmith1996b}.  On the other hand, our proposed
backdoors generate semi-primes with factors without common shared bits and do
not require Coppersmith's algorithm. Therefore, they are difficult to be detected and
are much more efficient when applied to semi-prime having large size, such as
those used in the currently used RSA public keys.

\section{\SSB: a backdoor embedded in a single semi-prime}
\label{s:ssb}

In this section we introduce \SSB\ (Single Semi-prime Backdoor), our proposal
for a backdoor encoded in the value of a semi-prime $N$.  We first describe the
vulnerability and how the semi-prime is generated; then, we describe the
procedure to efficiently factor it, provided that the corresponding ``escrow
key'' is known. Finally, we analyze the theoretical and practical
``efficiency'' of the backdoor.

\subsection{Generation of a vulnerable semi-prime}
\label{ss:ssbgen}

The first step of the generation of a vulnerable semi-prime is choosing a
``escrow key'', namely a secret value that must be known in order to detect and
exploit the backdoor. The escrow key is a prime $T$ of size slightly smaller
than the size of the primes in the semi-prime. Thus, if $\alpha$ is the
reference bit length of the primes (e.g., $\alpha=512$ for RSA-1024), then
$\bitsize{T}\simeq \alpha-c$, where typically $4\leq c\leq 10$ for $\alpha\leq
2048$; a good value for $\alpha=512$, $1024$, and $2048$ appears to be $c=7$.
The backdoor designer must also choose the value of a constant $K$, which must
be related to the value of $\alpha$, as discussed later; in particular, we will
show that a good practical choice may be $K\approx\alpha$.

\begin{figure}[t]
\begin{center}
\begin{minipage}[t]{\columnwidth}
\lstset{frame=tblr}
\begin{lstlisting}
GetCorrelPrimes:
  Input: $\alpha$, $T$, $K$
  Output: $p$, $k$

  while true
    generate random primes $q$ and $r$
        of size $\alpha$
    for $k$ := $2$ to $K$:
        $p$ := $r+(k*q-r)\bmod T$
        if $p$ is prime: return $p$, $q$
    end for
  end while
\end{lstlisting}
\end{minipage}
\end{center}
\caption{Generation of a vulnerable semi-prime with escrow key $T$}
\label{fig:ssbgen}
\end{figure}

In order to create a vulnerable semi-prime, two distinct primes $p$ and $q$,
each of them having bit length roughly $\alpha$, must be generated. The backdoor
exists whenever the following condition holds:
\begin{itemize}
        \item[(H0)]
                There exists a positive integer $k$ with $1<k\leq K$ such that
                $\cmod{p}{k\,q}{T}$.
\end{itemize}

The algorithm in Figure~\ref{fig:ssbgen} can be used to generate the two primes
$p$ and $q$ satisfying the condition H0 above. It is based on Dirichlet's
theorem stating that there are infinitely many primes of the form $a+b\,c$ if
$\gcd(a,b)=1$ (consider $b=T$ and $a=k\cdot(q\omod T)$).  The semi-prime is
then computed as $N=p\,q$.

\subsection{Recovering procedure}

The key idea of \SSB, and also the proof that it works as expected, is
its recovering procedure. Formally, the factors of $N$ can be
efficiently recovered by knowing in advance only the semi-prime $N$
and the escrow key $T$. The values of the parameters $\alpha$, $K$, and
$c$ may affect the running time of the recovering procedure, however there is
no need to know them to recover the factors.

The recovering procedure can be split in three phases:

\begin{enumerate}
    \item Recovering ``low-level'' coefficient.
    \item Recovering ``high-level'' coefficients.
    \item Recovering the factors.
\end{enumerate}

Generally speaking, in a practical implementation of the recovering procedure
it might be convenient to interleave the executions of these three phases.
However, we discuss the phases independently to simplify the description of the
whole procedure.

\ifrunexample\begin{runexample}\baselineskip=14pt
This ``running example'' is useful to understand the description of the
recovering procedure. Let $\alpha=128$, $c=5$, $K=30$.  We pick as a random
secret the 123-bit prime $T=6451117418610792529759522664972769997$. Then we
pick as vulnerable semi-prime
$N=54577680260424665710663143106120874652519112194523277824721618245793829954991$.
(of bit length 255).
\end{runexample}\fi

\subsubsection{Recovering ``low-level'' coefficients}

At the beginning we only know $N$ and $T$. The equation $N=p\,q$ and the equation
in condition~H0 imply:
\begin{equation}
        \cmod{N\omod T}{(p\omod T)\cdot(q\omod T)}{T}
\end{equation}
\begin{equation}
        \cmod{p\omod T}{(k\,q)\omod T}{T}
        \label{e:peqkq}
\end{equation}
By combining them we get:
\begin{equation}
        \cmod{N\omod T}{(k\,q^2)\omod T}{T}
\end{equation}

Because $k\in\left[2,K\right]$, where $K$ is a reasonably small constant, we
can exhaustively test every possible value for $k$ and discard any value for
which $N\cdot k^{-1}$ in the Galois field GF($T$) is a quadratic non-residue,
that is, discard any value $k$ such that for all integers
$\gamma\in\left[0,T\right)$, $\ncmod{(N\omod T)\,(k^{-1}\omod
T)}{\gamma^2}{T}$. Here $k^{-1}$ denotes the value in GF($T$) such that
$\cmod{k\cdot k^{-1}}{1}{T}$.

At the end of this phase we have a list containing
candidate values for the ``low-level'' coefficient $k$ and the corresponding quadratic
residue $\gamma^2$ in GF($T$). The correct value of $k$ yields $\cmod{\gamma^2}{q^2}{T}$.

\ifrunexample\begin{runexample}\baselineskip=14pt
    There are 14 values for $k\in[2,30]$ that yield a quadratic residue in
    GF($T$). They are:
    $3,4,9,10,12,13,14,16,19,22,23,25,27,30$.
\end{runexample}\fi

\subsubsection{Recovering ``high-level'' coefficients}

Let us assume that we start this phase by knowing $N$, $T$, $k$,
and $q^2\omod T$. Actually, we execute this phase for any candidate in the list built in
the previous phase and discard any candidate as soon as it yields inconsistent results. 

As first step, we compute the square root of $\gamma^2=q^2\omod T$ in
GF($T$), that is, we find the values whose square is congruent to $\gamma^2$
modulo $T$, typically by means of the Tonelli-Shanks algorithm
\cite{tonelli1891,shanks73}.
Because in general any square root has two distinct values in GF($T$),
we get two possible values $\gamma_1$ and $\gamma_2$ for $q\omod T$,
where $\cmod{\gamma_1}{T-\gamma_2}{T}$. In the following, let $\gamma$ be
either $\gamma_1$ or $\gamma_2$; we have to perform this phase with both
values and discard the one that yields inconsistent results.

We can easily compute the value $p\omod T$ from equation~\eqref{e:peqkq}, so
we may now assume to know the values $N$, $T$, $q\omod T$, and $p\omod T$.

The semi-prime $N$ can be written as:
\begin{equation}
    p\,q=\left(\pi\,T+(p\omod T)\right)\cdot
            \left(\nu\,T+(q\omod T)\right)\text{,}
\end{equation}
\noindent that is, if $\delta=\left( N - (p\omod T)\,(q\omod T)\right) / T$:
\begin{equation}
    \label{e:cimod}
    \delta = \pi\,\nu\,T +
        \pi\,(q\omod T)+\nu\,(p\omod T)\text{.}
\end{equation}

From the last equation we easily get the following bounds:
\begin{equation}
    \label{e:upperbound}
    \pi\,\nu \leq \left\lceil {N}/{T^2} \right\rceil
\end{equation}
\begin{equation}
    \label{e:lowerbound}
    (\pi+1)\,(\nu+1) \geq \left\lfloor {N}/{T^2}\right\rfloor
\end{equation}

Therefore,
$\bitsize{\pi}+\bitsize{\nu}\simeq\bitsize{\pi\,\nu}\simeq\bitsize{N/T^2}\simeq
2\alpha-2(\alpha-c)=2c$.  Because by construction $c$ is a small constant, we
can adopt a brute force approach to discover the missing ``high-level''
coefficients $\pi$ and $\nu$.  The brute force search guesses the value of the
sum $\pi+\nu$, starting from the lower bound $\left\lfloor\sqrt{2\,(\left\lfloor
N/T^2\right\rfloor-1)}\right\rfloor$ (from equation~\eqref{e:lowerbound}) and
ending at the upper bound $\left\lceil N/T^2 \right\rceil\approx 2^{2c}$ (from
equation~\eqref{e:upperbound}).

For any candidate value of the sum $\pi+\nu$, let us transform
equation~\eqref{e:cimod} by introducing an unknown $x=\pi$,
$C=\pi+\nu=x+\nu$, $a=q\omod T$, $b=p\omod T$:
\[
    x\,(C-x)\,T+a\,x+b\,(C-x)=\delta\text{,}
\]
that is,
\[
    T\,x^2+(b-a-C\,T)\,x+\delta-b\,C=0.
\]
Because we are looking for integer solutions for $x$ and $C-x$,  the brute
force attack just try all values for $C$, in increasing order, and immediately
discard any value such that
\[ \Delta=(b-a-C\,T)^2-4\,T\,(\delta-b\,C) \]
is not a square. If the value of $C$ survives, the solutions
\[ \left( C\,T+a-b\pm\sqrt{\Delta}\right) / (2\,T) \]
are computed; if either one of the solutions is an integral number, the
pair $(x,C-x)=(\pi,\nu)$ is recorded as a candidate solution.

\subsubsection{Recovering the factors}

When this phase starts, we know $N$, $T$, $p\omod T$, $q\omod T$, and
a list of candidate solutions $(\pi,\nu)$.

For any candidate solution $(\pi,\nu)$, we compute the corresponding
\[ p=\pi\,T+(p\omod T)\quad\text{and}\quad q=\nu\,T+(q\omod T)
\text{,}\]
then we simply verify whether $p\cdot q=N$.  One of the candidate solutions
certainly yields a factorization of the semi-prime.

\ifrunexample\begin{runexample}\baselineskip=14pt
Finally, we get:
\[ \begin{array}{lllll}
    p &=& \pi\,T+(p\omod T) &=& 313801445905602285635531222640499824151\\
    q &=& \nu\,T+(q\omod T) &=& 173924247235121781823208735516135244841\\
\end{array} \]
and we verify that $p\cdot q =$
\[ = 54577680260424665710663143106120874652519112194523277824721618245793829954991\]
$=N$.
\end{runexample}\fi

\subsection{Analysis}

We briefly describe here the time complexity of the \SSB's recovering
procedure. As explained in the previous subsection, the procedure starts by
recovering the ``low-level'' coefficients by means of an exhaustive search
among $O(K)$ possible values for $k$. For every candidate value we must execute
some operations in GF($T$) whose cost is in $O((\log T)^2)=O(\alpha^2)$, and
also the Tonelli-Shanks algorithm to determine if a value $<T$ is a quadratic
residue, which costs $O((\log T)^3)=O(\alpha^3)$ \cite{bernstein01}. The list
of candidate values for $k$ has expected length $K/2$, because in a finite
field with an odd number of elements any quadratic residue has two square
roots, thus half of the elements of the field are not square of another
element.  Therefore, the ``high-level'' coefficients recovery phase is executed
on $O(K)$ candidate values for $k$ and includes an exhaustive search in an
interval of size $O(2^{2c})$; in every iteration we execute a few integer
operations on values of bit length $\approx 2(\alpha+c)$; hence, every execution
of this phase has a cost in $O(2^{2c}\,(\alpha+c)^2)$.  Finally, the cost of
every execution of the third phase is dominated by two multiplications of
values of bit length $\approx \alpha-c$, hence it is in $O(\alpha^2)$.  Summing
all up, the worst-case cost of the whole recovering procedure is in
$O(K\,(\alpha+c)^3\,2^{2c})$.

The values of the parameters $K$ and $c$ are chosen by the backdoor designer.
We would expect that larger values of $K$ and $c$ yield smaller running
times for the search algorithm in Figure~\ref{fig:ssbgen} and longer running
times for the recovery procedure; this intuition is confirmed by the experiments.
Anyway, the value of $c$ cannot be made too large, or it would be possible to
discover the backdoor by just guessing the design key $T$ of bit length
$\bitsize{T}=\alpha-c$.  By letting $K\in O(\alpha)$ and $c\in O(\log\alpha)$,
for instance $K\approx \alpha$ and $c=7$ as suggested in
subsection~\ref{ss:ssbgen}, we obtain a running time for the recovery procedure
in $O(\alpha^{4})$, that is, polynomial in the size of the semi-prime.

\subsubsection{Experimental results}

In order to confirm that the backdoor works as expected and to assess the
execution times with respect to the designer's parameters, we implemented \SSB\
in SageMath~\cite{sagemath} and performed extensive tests.\footnote{The code is
open-source and available at
\url{https://gitlab.com/cesati/ssb-and-tsb-backdoors.git}.}

In particular, we considered three values for $\alpha$: $512$ (the size of
factors for RSA-1024), $1024$ (RSA-2048), and $2048$ (RSA-4096). All tests have
been performed by choosing $c=7$. This means that the escrow keys have sizes
$505$, $1017$, and $2041$, respectively.  The value of $c$ is so small that
detecting the existence of the backdoor by simply guessing the value of the
escrow key does not appear to be significantly easier than guessing one of the
factors of the corresponding semi-primes. Every test trial involves
choosing a value for the parameter $K$, generating a escrow key $T$ and a
vulnerable semi-prime, then recovering the factors of the semi-prime by just
using the values of the semi-prime and the escrow key. We basically executed
the tests by varying the parameter $K$ so as to determine a value yielding both
fast generations of vulnerable semi-primes and reasonably quick recovery of the
factors.

\begin{table}
    \begin{filecontents*}{ssb512.csv}
K,GEN,GSD,REC,RSD
100,4.9,4.8,2.6,1.8
500,1.7,0.6,7.8,6.9
1000,1.5,0.2,12.2,9.6
1500,1.5,0.1,18.0,17.1
2000,1.4,0.1,10.0,9.0
2500,1.6,0.1,15.3,11.2
3000,1.7,0.1,15.7,15.7
3500,1.7,0.1,26.1,30.3
4000,1.7,0.1,14.6,18.8
4500,1.6,0.2,19.0,16.6
5000,1.6,0.3,24.8,19.0
\end{filecontents*}
\pgfplotstabletypeset[
      multicolumn names, % allows to have multicolumn names
      col sep=comma, % the separator in our .csv file
      display columns/0/.style={
        column name={$K$}, % name of first column
        column type={r}},
      display columns/1/.style={
        column name={avg.},
        column type=r,string type},
      display columns/2/.style={
        column name={st.dev.},
        column type=r,string type},
      display columns/3/.style={
        column name={avg.},
        column type=r,string type},
      display columns/4/.style={
        column name={st.dev.},
        column type=r,string type},
      every head row/.style={
        before row={\toprule
        $\alpha{=}512$&\multicolumn{2}{c}{Generation}&\multicolumn{2}{c}{Recovering}\\
        },
        after row=\midrule
      },
      every last row/.style={after row=\bottomrule} % rule at bottom
    ]{ssb512.csv} % filename/path to file
    \par\medskip
    \begin{filecontents*}{ssb1024.csv}
K,GEN,GSD,REC,RSD
100,51.0,60.3,5.4,1.9
500,17.9,7.4,18.3,13.5
1000,11.6,3.2,34.9,30.7
1500,11.2,1.6,33.9,37.8
2000,11.5,1.8,28.6,23.0
2500,10.6,1.3,61.2,60.0
3000,11.4,1.7,45.4,63.7
3500,10.8,0.9,72.0,63.4
4000,10.7,1.5,56.0,49.0
4500,10.8,1.0,42.7,41.9
5000,11.1,1.2,44.4,38.4
\end{filecontents*}
\pgfplotstabletypeset[
      multicolumn names, % allows to have multicolumn names
      col sep=comma, % the separator in our .csv file
      display columns/0/.style={
        column name={$K$}, % name of first column
        column type={r}},
      display columns/1/.style={
        column name={avg.},
        column type=r,string type},
      display columns/2/.style={
        column name={st.dev.},
        column type=r,string type},
      display columns/3/.style={
        column name={avg.},
        column type=r,string type},
      display columns/4/.style={
        column name={st.dev.},
        column type=r,string type},
      every head row/.style={
        before row={\toprule
      $\alpha{=}1024$&\multicolumn{2}{c}{Generation}&\multicolumn{2}{c}{Recovering}\\
        },
        after row=\midrule
      },
      every last row/.style={after row=\bottomrule} % rule at bottom
    ]{ssb1024.csv} % filename/path to file
    \par\medskip
    \begin{filecontents*}{ssb2048.csv}
K,GEN,GSD,REC,RSD
100,466.4,375.1,27.1,6.9
500,214.0,134.7,47.4,18.6
1000,151.2,61.3,71.6,37.4
1500,122.3,35.0,101.2,70.0
2000,102.7,20.3,95.8,51.7
2500,112.7,25.4,113.2,84.8
3000,107.6,23.0,130.7,84.1
3500,99.5,22.6,95.5,54.8
4000,90.4,9.1,143.1,104.4
4500,91.5,13.5,152.8,136.6
5000,97.3,16.9,94.5,91.5
\end{filecontents*}
\pgfplotstabletypeset[
      multicolumn names, % allows to have multicolumn names
      col sep=comma, % the separator in our .csv file
      display columns/0/.style={
        column name={$K$},
        column type={r}},
      display columns/1/.style={
        column name={avg.},
        column type=r,string type},
      display columns/2/.style={
        column name={st.dev.},
        column type=r,string type},
      display columns/3/.style={
        column name={avg.},
        column type=r,string type},
      display columns/4/.style={
        column name={st.dev.},
        column type=r,string type},
      every head row/.style={
        before row={\toprule
        \multicolumn{3}{l}{$\alpha{=}2048$\hskip3.5ex Generation}&\multicolumn{2}{c}{Recovering}\\
        },
        after row=\midrule
      },
      every last row/.style={after row=\bottomrule} % rule at bottom
    ]{ssb2048.csv} % filename/path to file

    \caption{\SSB: running times (in seconds, average and standard
    deviations on 20 trials) for $\alpha=512, 1024, 2048$.} 
    \label{t:ssb}
\end{table}

\begin{figure}
\includegraphics[width=\columnwidth]{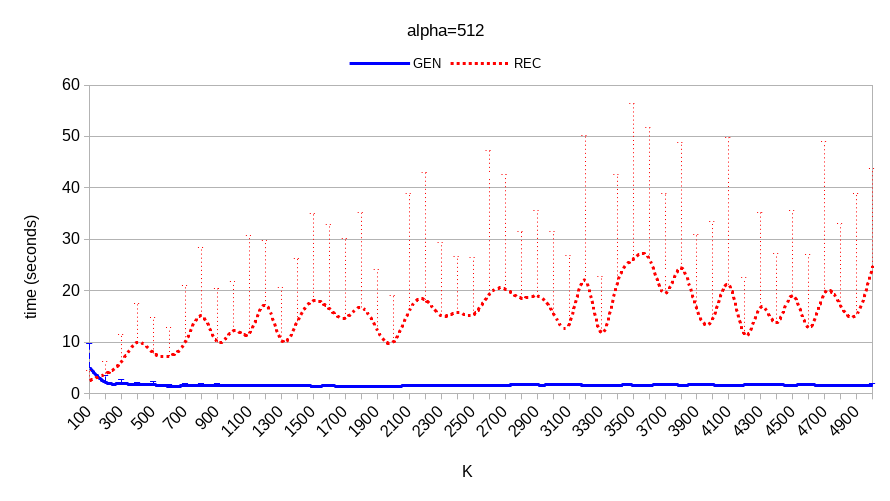}
\par\medskip
\includegraphics[width=\columnwidth]{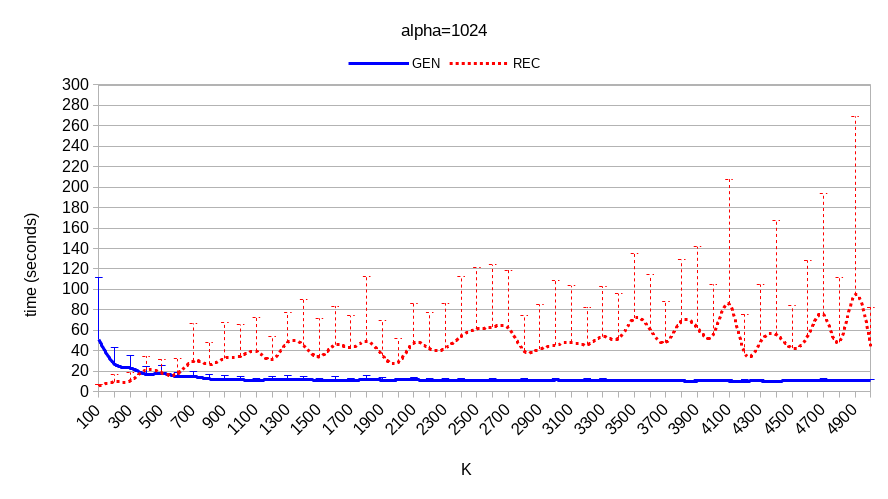}
\par\medskip
\includegraphics[width=\columnwidth]{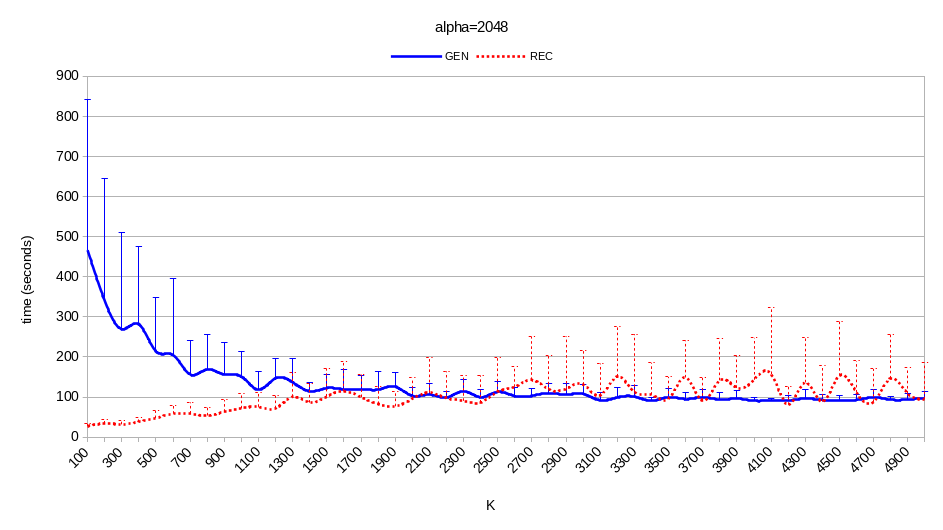}
\caption{\SSB: average running times for $\alpha=512, 1024, 2048$ (20 trials
for each value of $K$).  Magnitudes of the standard deviations are shown as
vertical bars.}
\label{f:ssb}
\end{figure}

The tests have been executed on three computational nodes with 16 physical
cores Intel Xeon E5-2620 v4 running at 2.1~GHz with 64~GiB of RAM. The nodes
are based on the Slackware~14.2 software distribution with a Linux kernel
version~5.4.78 and SageMath version~9.1. All tests have properly recovered the
factors of the vulnerable semi-primes. Each value of $K\in\set{100\cdot i\,|\,
i=1,\ldots, 50}$ has been tested 20 times. The SageMath code is sequential,
that is, each test trial runs on a single computation core. We report in
Table~\ref{t:ssb} and Figure~\ref{f:ssb} averages and standard deviations of
the running times.

The experimental results confirm that the value of $K$ is crucial in determining
both the time required to generate a vulnerable semi-prime and the time
required to recover the factors. Even if the code has not been optimized at
all, the recovery time is reasonably small for all tested values of~$K$, hence
\SSB\ is a practically effective backdoor. However, generation time is also
very important whenever the backdoor mechanism has to be hidden in hardware
devices or software programs that are supposed to yield robust, legit
semi-primes. While in general larger values of~$K$ are associated to smaller
generation times, there seems to be a threshold value for $K$ above which the
generation times are essentially constants and near the minimum observed value.
From the values shown in Table~\ref{t:ssb} and Figure~\ref{f:ssb} we may safely
set $K$ to values near $500$, $1000$, and $2000$ for $\alpha=512$, $1024$, and
$2048$, respectively, that is, $K\approx\alpha$.

% GPG RSA key generation times on the nodes:
% with /dev/urandom c(1,9):
% alpha =  512 (RSA-1024) : avg 0.04 s, st.dev. 0.02 s
% alpha = 1024 (RSA-2048) : avg 0.4 s, st.dev. 0.1 s
% alpha = 2048 (RSA-4096) : avg 4.3 s, st.dev. 2.5 s
% with /dev/random c(1,8):
% alpha =  512 (RSA-1024) : ???
% alpha = 1024 (RSA-2048) : ???
% alpha = 2048 (RSA-4096) : ???

\section{\TSB: a backdoor embedded in a pair of semi-primes}
\label{s:tsb}

In this section we describe \TSB\ (Twin Semi-prime Backdoor), our proposal for
a backdoor embedded in the values of a pair of semi-primes $N_1$ and~$N_2$.
These semi-primes are typically to be generated on the same device but can be
used independently.  For instance, the two semi-primes might be used in two
different RSA keys.\footnote{It is not hard to justify the generation of two
different RSA keys. For instance, the user might be told that one RSA key is
for business or work usage and the other one is for personal usage.}
Alternatively, one of the semi-prime can be used to build a RSA key while the
other one would be separately stored as an escrow key for the RSA key.

We first describe how the two semi-primes are generated. Then we describe the
procedure to efficiently factor both semi-primes, provided that the
corresponding ``designer key'' is known.

\subsection{Generation of the vulnerable pair of semi-primes}
\label{ss:tsbgen}

The first step of the generation of a vulnerable pair is choosing a ``designer
key'', namely a secret value that must be known in order to detect and exploit
the backdoor. The designer key is a prime $T$ of size slightly smaller than the
size of the primes in each semi-prime. Thus, if $\alpha$ is the reference
bit length of the primes (e.g., $\alpha=512$ for RSA-1024), then
$\bitsize{T}\simeq \alpha-c$, where typically $4\leq c\leq 10$ for $\alpha\leq 2048$;
a good value for $\alpha=512$, $1024$, and $2048$ appears to be $c=7$.
The backdoor designer must also choose the values of two constants $K$ and $B$.
The value of $K$ is related to the value of $\alpha$, as discussed later; typically,
we may set $K\approx\alpha/5$, e.g., $K=100$, $200$, and $400$ for $\alpha=512$,
$1024$, and $2048$, respectively.  The constant $B<T$ acts as a detection
threshold, so any value for $B$ such that $\bitsize{B}\simeq \alpha-2c$ is
valid.

In order to create a vulnerable pair, four distinct primes $p_1$, $q_1$, $p_2$,
$q_2$, each of them having bit length roughly $\alpha$, must be generated. The
backdoor exists whenever the following conditions hold:

\begin{itemize}
        \item[(H1)]
                There exists a positive integer $h$ with $1<h\leq K$ such that
                $\cmod{q_2}{h^2\,q_1}{T}$.
%       \begin{equation}
%           \label{eq:qcong}
%           \cmod{q_2}{h^2\,q_1}{T}\text{.}
%       \end{equation}
\item[(H2)]\label{eq:p1cong}
        There exists a positive integer $k_1$ with $1<k_1\leq K$ such
        that $\cmod{p_1}{h\,k_1\,q_2}{T}$.
%       \begin{equation}
%           \label{eq:p1cong}
%           \cmod{p_1}{h\,k_1\,q_2}{T}\text{.}
%       \end{equation}
\item[(H3)]\label{eq:p2cong}
        There exists a positive integer $k_2$ with $1<k_2\leq K$ such
        that  $\cmod{p_2}{k_2\,q_1}{T}$.
%       \begin{equation}
%           \label{eq:p2cong}
%           \cmod{p_2}{k_2\,q_1}{T}\text{.}
%       \end{equation}
\item[(H4)]\label{eq:gcd}
        The integers $h$, $k_1$, and $k_2$ are all coprime, that is,
                $\gcd(h,k_1)=\gcd(h,k_2)=\gcd(k_1,k_2)=1$.
%        \begin{equation}
%            \label{eq:gcd}
%            \gcd(h,k_1)=\gcd(h,k_2)=\gcd(k_1,k_2)=1\text{.}
%        \end{equation}
\item[(H5)]\label{eq:ncong}
        $k_2$ is not a divisor of $h\,k_1$ modulo $T$, that is, $\ncmod{h\,k_1}{k_2}{T}$.
%       \begin{equation}
%           \label{eq:ncong}
%           \ncmod{h\,k_1}{k_2}{T}\text{.}
%       \end{equation}
\item[(H6)]\label{eq:diseg}
        Finally, $(h\,q_1)^2\omod T>B$.
%        \begin{equation}
%            \label{eq:diseg}
%            (h\,q_1)^2\omod T>B\text{.}
%        \end{equation}
\end{itemize}

The algorithm in Figure~\ref{fig:generation} can be used to generate the four primes $p_1$,
$q_1$, $p_2$, and $q_2$ satisfying the conditions H1--H6 above. Once more, the algorithm
is implicitly based on Dirichlet's theorem stating that there are infinitely many primes
of the form $a+b\,c$ when $\gcd(a,b)=1$.

\begin{figure}[h]
\begin{minipage}[t]{\columnwidth}
\lstset{frame=tlrb}
\begin{lstlisting}
GeneratePair:
  Input: $\alpha$, $c$, $K$, $T$
  Output: $p_1$, $q_1$, $p_2$, $q_2$

do
  generate random primes $q_1$, $p$ of size $\alpha$
  for $h$ := $2$ to $K$
      $q_2$ := $p+((h^2\,q_1-p)\bmod T)$
      if $q_2$ is prime then
         break for loop
      end if
  end for
while $q_2$ is not prime

do
  $p_1$, $k_1$ := GetCorrelPrime($\alpha$,$q_2$,$h$,$T$,$K$,$c$)
while gcd($k_1$,$h$) $== 1$

do
  $p_2$, $k_2$ := GetCorrelPrime($\alpha$,$q_1$,$1$,$T$,$K$,$c$)
while gcd($k_1$,$k_2$) == 1 or gcd($k_2$,$h$) == 1
return $p_1$, $q_1$, $p_2$, $q_2$

GetCorrelPrime:
  Input: $\alpha$, $q$, $j$, $T$, $K$, $c$
  Output: $p$, $k$

  while true
    $k$ := random value between $2$ and $K$
    $t_1$ := $(k\,j\,q)\bmod T$
    for $p$ := $t_1$ + $2^{c-3}$ to $t_1$ + $2^{2\,c-2}$ 
    if $\cmod{p}{t_1}{T}$ and $p$ is prime then
        return $p$, $k$
    end if
  end while
\end{lstlisting}
\end{minipage}
\caption{Generation of a vulnerable pair of semi-primes with designer key $T$}
\label{fig:generation}
\end{figure}

Finally, the semi-primes are computed as $N_1=p_1\,q_1$ and $N_2=p_2\,q_2$.
Observe that $N_1$ and $N_2$ are coprime, because all factors are necessarily
different by construction.

\iffalse
\emph{TODO: we should also enforce that $p_i, q_i$ 
are safe primes, that is, that $\frac{p-1}{2}$ is a prime. There should be no
problem in that, however this has to be verified. The issue here is that we need
an efficient way to generate vulnerable semi-prime pairs: the naive GP program is
too slow and generates semi-primes that are not safe.}
\fi

\subsection{Recovering procedure}

The key idea of \TSB, and also the proof that it works as expected, is
its recovering procedure. Formally, the factors of $N_1$ and $N_2$ can be
efficiently recovered by knowing in advance only the pair of semi-primes $(N_1,
N_2)$ and the designer key $T$. The values of the parameters $\alpha$, $K$, and
$c$ may affect the running time of the recovering procedure, however there is
no need to know them to recover the factors.

The recovering procedure can be split in four phases:

\begin{enumerate}
    \item Recovering ``medium-level'' coefficients.
    \item Recovering ``low-level'' coefficients.
    \item Recovering ``high-level'' coefficients.
    \item Recovering the factors.
\end{enumerate}

Generally speaking, in a practical implementation of the recovering procedure
it might be convenient to interleave the executions of these three phases.
However, we discuss the phases independently to simplify the description of the
whole procedure.

\ifrunexample\begin{runexample}\baselineskip=14pt
This ``running example'' is useful to understand the description of the
recovering procedure. Let $\alpha=64$, $c=3$, $K=100$, $B=2^{57}$.  We pick as
a random secret the 61-bit prime $T=1350856093440009833$. Then we compute as
vulnerable semi-primes $N_1=199771249142689629600100193795300988277$ and
$N_2=330849388672597230630022641974377014199$ (both of bit length 128).
\end{runexample}\fi

\subsubsection{Recovering ``medium-level'' coefficients}

When starting the recovering procedure we assume to know the following data:
$N_1$, $N_2$, and the ``secret'' prime $T$.

Equations in conditions H1, H2, and H3 enforce the following congruences of
$N_1$ and $N_2$ modulo $T$:
\begin{equation}
    \label{eq:n1cong}
    \cmod{N_1}{p_1\,q_1\modequiv{T}h\,k_1\,q_2\,q_1\modequiv{T}
        h^3\,k_1\,q_1^2}{T}
\end{equation}
\begin{equation}
    \label{eq:n2cong}
    \cmod{N_2}{p_2\,q_2\modequiv{T}k_2\,q_1\,q_2\modequiv{T}
        h^2\,k_2\,q_1^2}{T}
\end{equation}

It turns out that $N_1$ and $N_2$ are congruent modulo $T$ to two values that
have a big common factor, $h^2\,q_1^2$. However, the Euclidean algorithm on
$N_1\omod T$ and $N_2\omod T$ does not really help here:
\[ \begin{array}{l}
        \gcd(N_1\omod T, N_2\omod T) = \\[6pt]
   \gcd((h^3\,k_1\,q_1^2)\omod T,
   (h^2\,k_2\,q_1^2)\omod T)\text{.} \end{array} \]
The point is that the greatest common divisor is relative to the lifted images
of the products in the Galois field GF($T$), and it is not related to the
greatest common divisor of the products $h^3\,k_1\,q_1^2$ and $h^2\,k_2\,q_1^2$
in $\mathbb{Z}$.

\ifrunexample\begin{runexample}\baselineskip=14pt
In our running example:
$$\gcd(N_1\omod T, N_2\omod T)=\gcd(337598081507736831,75151731210637471)=1\text{.}$$
\end{runexample}\fi

To overcome this problem, observe that equations \eqref{eq:n1cong} and
\eqref{eq:n2cong} also imply the following ones:
\begin{equation}
    \cmod{N_1\omod T}{(h\,k_1)\cdot \left[(h^2\,q_1^2)\omod T\right]}{T}
\end{equation}
\begin{equation}
     \cmod{N_2\omod T}{k_2\cdot \left[(h^2\,q_1^2)\omod T\right]}{T}
\end{equation}
\noindent and therefore there exist two integers $\tilde{k_1}$, 
$\tilde{k_2}$ such that
\begin{equation}
    \label{eq:mediumcong1}
    (N_1\omod T)+\tilde{k_1}\cdot T=(h\,k_1)\cdot \left[(h^2\,q_1^2)\omod
    T\right]
\end{equation}
\begin{equation}
    \label{eq:mediumcong2}
    (N_2\omod T)+\tilde{k_2}\cdot T=k_2\cdot \left[(h^2\,q_1^2)\omod T\right]
\end{equation}

From the last two equations we derive:
\begin{equation}
    \label{eq:gcdmedium}
    \begin{array}{r}
    \gcd((N_1\omod T)+\tilde{k_1}\cdot T,\,\\[6pt]
         (N_2\omod T)+\tilde{k_2}\cdot T)\\[6pt]
         =\left[(h^2\,q_1^2)\omod T\right] 
    \end{array}
\end{equation}

Observe that dropping $N_1\omod T$ from equation~\eqref{eq:mediumcong1} yields
\[ \tilde{k_1} \leq (h\,k_1)\cdot\frac{(h^2\,q_1^2)\omod T}{T} < K^2\text{.} \]
Similarly, from equation~\eqref{eq:mediumcong2}: \[ \tilde{k_2} \leq
k_2\cdot\frac{(h^2\,q_1^2)\omod T}{T}<K\text{.}\] Hence, the sizes of the
``medium'' coefficients $\tilde{k_1}$ and $\tilde{k_2}$ is so small that they
can be quickly recovered by a brute force approach as in
Figure~\ref{fig:mediumsearch}.

\begin{figure}[h]
\begin{center}
\begin{minipage}{\columnwidth}
\lstset{frame=tblr}
\begin{lstlisting}
RecoveryMedCoeff:
  Input: $N_1$, $N_2$, $T$
  Output: a list of pairs $(\tilde{k_1}$, $\tilde{k_2})$

for s := 0 to $\infty$
  for $\tilde{k_1}$ := 0 to s
    $\tilde{k_2}$ := s - $\tilde{k_1}$
    gg := gcd($\tilde{k_1}\cdot T+N_1\omod T$,
              $\tilde{k_2}\cdot T+N_2\omod T$)
    if ($B<$gg$<T$) and
       ($\cmod{\text{gg}}{\gamma^2}{T}$ for some $\gamma$) then
      add $(\tilde{k_1}$, $\tilde{k_2})$ to the list of coeff.
    end if
  end for
end for
\end{lstlisting}
\end{minipage}
\end{center}
\caption{Brute-force search of the medium-level coefficients}
\label{fig:mediumsearch}
\end{figure}

It is possible to recognize the proper values of $\tilde{k_1}$ and $\tilde{k_2}$
because the size of $\left[(h^2\,q_1^2)\omod T\right]$ produced by the gcd with
the right values is usually much higher than the average value resulting from a
gcd with random wrong values.  In fact, by condition~H6, $(h^2\,q_1^2)\omod
T>B$; hence we select any candidate pair of medium-level coefficients
$(\tilde{k_1}, \tilde{k_2})$ for which the greatest common divisor in equation
\eqref{eq:gcdmedium} is between $B$ and $T$. Moreover, the value returned by
the Euclidean algorithm with the right values must be a square in the Galois
field GF($T$), hence we may use this condition to filter some false positives.
In all our test cases, the first value found by this brute force procedure
yields a proper factorization result.

\ifrunexample\begin{runexample}\baselineskip=14pt
    There are only two possible pairs $(\tilde{k_1},\tilde{k_2}) \in
    [0,100^2]\times [0,100]$ that yield a greatest common divisor higher than
    $B=2^{57}$: $(671,10)$ and $(5277,79)$.  The gcd of the pair  $(671,10)$ is
    $196865400950880229$, which is the square of $10632559655363908$ modulo
    $T$.  The gcd of the pair  $(5277,79)$ is $1547721494390890062$, which is
    above $T$ and can be discarded.
\end{runexample}\fi

\subsubsection{Recovering ``low-level'' coefficients}

The previous phase might determine several candidate pairs of medium-level
coefficients, and the current phase must be applied to each of them.

Let us assume at this point to know the following data: $N_1$, $N_2$, $T$,
$\tilde{k_1}$, $\tilde{k_2}$, and the value $\gamma^2=\left[(h^2\,q_1^2)\omod
T\right]$ derived from equation~\eqref{eq:gcdmedium}.
The value of the ``low-level'' coefficient $k_2$ can be immediately computed
by using equation~\ref{eq:mediumcong2}:
\begin{equation}
    k_2 = \left( (N_2\omod T) + \tilde{k_2}\cdot T \right) / \gamma^2
    \text{,}
\end{equation}
or, assuming $K<T$, $k_2=(N_2\cdot (\gamma^2)^{-1})\omod T$ where
$\cmod{\gamma^2\cdot(\gamma^2)^{-1}}{1}{T}$.

On the other hand, by inverting equation~\ref{eq:mediumcong1} we get the value
of the product $h\,k_1$:
\begin{equation}
    (h\,k_1) = \left( (N_1\omod T) + \tilde{k_1}\cdot T \right) /
    \gamma^2\text{,}
\end{equation}
or, assuming $K<T$, $(h\,k_1)=(N_1\cdot (\gamma^2)^{-1})\omod T$.

Since both $h$ and $k_1$ are not greater than $K$, their product is below $K^2$.
Moreover, by condition~H4, $\gcd(h, k_1)=1$. Because the number of
multiplicative partitions of this product does not exceed $K^2$ 
\cite{canfield1983,hughes1983}, we may exhaustively generate all possible
candidate pairs $(h,k_1)$ and apply the forthcoming phases to each of them.
When these phases are performed on the true pair $(h,k_1)$, a proper
factorization of $N_1$ and $N_2$ is computed.

\ifrunexample\begin{runexample}\baselineskip=14pt
    Two exact integer divisions yield $k_2=69$ and $(h\,k_1)=4606=2\cdot
    7^2\cdot 47$.  Therefore, there are six possible pairs $(h, k_1)$,
    corresponding to the non-trivial subsets of the three values $2$, $7^2$, and
    $47$: $(2,2303)$, $(47,98)$, $(49,94)$, $(94,49)$, $(98,47)$, and $(2303,2)$.
\end{runexample}\fi

\subsubsection{Recovering ``high-level'' coefficients}

Let us assume that we start this phase by knowing the following data:
$N_1$, $N_2$, $T$, $h$, $k_1$, $k_2$, and $\gamma^2$. 

As first step, we compute the square root of $\gamma^2=(h^2\,q_1^2)\omod T$ in
GF($T$), that is, we find the values whose square is congruent to $\gamma^2$
modulo $T$, typically by means of the Tonelli-Shanks algorithm
\cite{tonelli1891,shanks73}.
Because in general any square root has two distinct values in GF($T$),
we get two possible values $\gamma_1$ and $\gamma_2$ for $(h\,q_1)\omod T$,
where $\cmod{\gamma_1}{T-\gamma_2}{T}$. In the following, let $\gamma$ be
either $\gamma_1$ or $\gamma_2$; we have to perform this phase with both
values and discard the one that yields inconsistent results.

We can now compute the value $q_1\omod T$, because $\gamma=(h\,q_1)\omod T$
means:
\begin{equation}
    q_1\omod T=(\gamma\,h^{-1})\omod T
\end{equation}
\noindent where obviously $h^{-1}$ is computed in GF($T$), that is, 
$\cmod{h\,h^{-1}}{1}{T}$.

\iffalse
Alternatively, we could compute $q_1\omod T$ from either one of
equations~\eqref{eq:n1cong} and~\eqref{eq:n2cong}:

\begin{equation}
    \cmod{q_1}{\sqrt{N_1\cdot (h^3\,k_1)^{-1}}}{T}
\end{equation}

\begin{equation}
    \cmod{q_1}{\sqrt{N_2\cdot (h^2\,k_2)^{-1}}}{T}
\end{equation}
\fi

From equation in condition~H1 we can now infer the value $q_2\omod T$, because:
\begin{equation}
    q_2\omod T=\left((q_1\omod T)\cdot h^2\right)\omod T
\end{equation}

Also, from equations in conditions~H2 and~H3 we compute
$p_1\omod T$ and $p_2\omod T$:
\begin{equation}\begin{array}{c}
    p_1\omod T=\left(h\,k_1\,(q_2\omod T)\right)\omod T,\\[6pt]
    p_2\omod T=\left(k_2\,(q_1\omod T)\right)\omod T\text{.}
    \end{array}
\end{equation}

At this point we know the values $N_1$, $N_2$, $T$, $q_1\omod T$, $q_2\omod T$,
$p_1\omod T$, and $p_2\omod T$.

The semi-prime $N_i$ ($i\in\set{1,2}$) can be written as:
\begin{equation}\begin{array}{l}
    N_i=p_i\,q_i = \\[6pt]
        \left(\pi_i\,T+(p_i\omod T)\right)\cdot
            \left(\nu_i\,T+(q_i\omod T)\right)\text{,}
\end{array}\end{equation}
\noindent that is, if $\delta_i=\left( N_i - (p_i\omod T)\,(q_i\omod T)\right) / T$:
\begin{equation}
    \label{eq:cimod}
    \delta_i = \pi_i\,\nu_i\,T +
        \pi_i\,(q_i\omod T)+\nu_i\,(p_i\omod T)\text{.}
\end{equation}

From the last equation we easily get the following bounds:
\begin{equation}
    \label{eq:upperbound}
    \pi_i\,\nu_i \leq \left\lceil {N_i}/{T^2} \right\rceil
\end{equation}
\begin{equation}
    \label{eq:lowerbound}
    (\pi_i+1)\,(\nu_i+1) \geq \left\lfloor {N_i}/{T^2}\right\rfloor
\end{equation}

Therefore, $\bitsize{\pi_i}+\bitsize{\nu_i}\simeq 2\alpha-2(\alpha-c)=2c$.
Because by construction $c$ is a small constant, we can adopt a brute force
approach to discover the missing ``high-level'' coefficients $\pi_i$ and
$\nu_i$.  The brute force search guesses the value of the sum
$\pi_i+\nu_i$, starting from the lower bound $\left\lfloor\sqrt{2(\left\lfloor
N_i/T^2\right\rfloor-1)}\right\rfloor$ (from equation~\eqref{eq:lowerbound})
and ending at the upper bound $\left\lceil N_i/T^2 \right\rceil\approx 2^{2c}$
(from equation~\eqref{eq:upperbound}).

For any candidate value of the sum $\pi_i+\nu_i$, let us transform
equation~\eqref{eq:cimod} by introducing an unknown $x=\pi_i$,
$C=\pi_i+\nu_i=x+\nu_i$, $a_i=q_i\omod T$, $b_i=p_i\omod T$:
\[
    x\,(C-x)\,T+a_i\,x+b_i\,(C-x)=\delta_i\text{,}
\]
that is,
\[
    T\,x^2+(b_i-a_i-C\,T)\,x+\delta_i-b_i\,C=0.
\]
Because we are looking for integer solutions for $x$ and $C-x$,  the brute
force attack just try all values for $C$, in increasing order, and immediately
discard any value such that
\[ \Delta=(b_i-a_i-C\,T)^2-4\,T\,(\delta_i-b_i\,C) \]
is not a square. If the value of $C$ survives, the solutions
\[ \left( C\,T+a_i-b_i\pm\sqrt{\Delta}\right) / (2\,T) \]
are computed; if either one of the solutions is an integral number, the
pair $(x,C-x)=(\pi_i,\nu_i)$ is recorded as a candidate solution.

\subsubsection{Recovering the factors}

When this phase starts, we know $N_i$, $T$, $p_i\omod T$, $q_i\omod T$, and
a list of candidate solutions $(\pi_i,\nu_i)$, for $i=1,2$. We work on every
semi-prime separately.

For any candidate solution $(\pi_i,\nu_i)$, we compute the corresponding
$p_i=\pi_i\,T+(p_i\omod T)$ and\linebreak $q_i=\nu_i\,T+(q_i\omod T)$, then we simply
verify whether $p_i\cdot q_i=N_i$.  One of the candidate solutions certainly
yields a factorization of the semi-prime.

\ifrunexample\begin{runexample}\baselineskip=14pt
Finally, we get:
\[ \begin{array}{lllll}
    p_1 &=& \pi_1\,T+(p_1\omod T) &=& 12258708750247312273\\
    q_1 &=& \nu_1\,T+(q_1\omod T) &=& 16296271753634685349\\
    p_2 &=& \pi_2\,T+(p_2\omod T) &=& 16740754379985226021\\
    q_2 &=& \nu_2\,T+(q_2\omod T) &=& 19763111097798043819\\
\end{array} \]

and we verify that 

\[ \begin{array}{lllll}
    p_1 \cdot q_1 &=& 199771249142689629600100193795300988277 &=& N_1\\
    p_2 \cdot q_2 &=& 330849388672597230630022641974377014199 &=& N_2\\
\end{array} \]
\end{runexample}\fi

\subsection{Analysis}
\label{ss:tsbanal}

We briefly describe here the time complexity of the \TSB's recovering
procedure. As already explained, the procedure starts by recovering the
``medium-level'' coefficients by means of an exhaustive search among $K^3$
possible values for the pair $(\tilde{k_1},\tilde{k_2})$. For every candidate
pair we must execute the Euclidean algorithm on values of bit length up to
$\approx \bitsize{\tilde{k_1}\,T}$, which costs
$O(\log(\tilde{k_1}\,T))=O(\bitsize{\tilde{k_1}}+\bitsize{T})=O(\log
K+\alpha-c)$.  We may also use the Tonelli-Shanks algorithm to determine if a
value $<T$ is a quadratic residue, which costs $O((\log T)^3)=O(\alpha^3)$
\cite{bernstein01}. The ``low-level'' coefficients recovery phase involves a
couple of integer divisions on values $\approx\tilde{k_1}\,T$, a factorization
of a value $<K^2$, and the generation of up to $K^2$ candidate pairs $(h,k_1)$;
hence, each execution of this recovery phase has a cost in
$O(\alpha^2+K^2)$. The ``high-level'' coefficients recovery phase includes
an exhaustive search in an interval of size $O(2^{2c})$; in every iteration we
execute a few integer operations on values of bit length $\approx 2(\alpha+c)$;
hence, every execution of this phase has a cost in $O(2^{2c}\,(\alpha+c)^2)$.
Finally, the cost of every execution of the fourth phase is dominated by four
multiplications of values of bit length $\approx \alpha-c$, hence it is in
$O(\alpha^2)$.  Summing all up, the worst-case cost of the whole recovering
procedure is in $O(K^5\,(\alpha+c)^2\,2^{2c})$.

The values of the parameters $K$ and $c$ are chosen by the backdoor designer.
It is easy to observe that larger values of $K$ and $c$ yield shorter running
time for the search algorithm in Figure~\ref{fig:generation} and longer running
time for the recovery procedure.  Anyway, the value of $c$ cannot be made too
large, or it would be possible to discover the vulnerability by just guessing
the design key $T$ of bit length $\bitsize{T}=\alpha-c$. On the other hand,
experimental results show that larger values of $c$ do not necessarily yield
shorter times for the generation phase. By letting $K\approx\alpha/5$ and
$c=7$, as suggested in subsection~\ref{ss:tsbgen}, we obtain a running time for
the recovery procedure in $O(\alpha^{7})$, that is, polynomial in the size
of the semi-primes.

\subsubsection{Experimental results}

In order to confirm that the backdoor works as expected and to assess the
execution times with respect to the designer's parameters, we implemented \TSB\
in SageMath~\cite{sagemath} and performed extensive tests.\footnote{The code is
open-source and available at
\url{https://gitlab.com/cesati/ssb-and-tsb-backdoors.git}.}

In particular, we considered three values for $\alpha$: $512$ (the size of
factors for RSA-1024), $1024$ (RSA-2048), and $2048$ (RSA-4096). All tests have
been performed by choosing $c=7$. This means that the designer keys have sizes
$505$, $1017$, and $2041$, respectively.  The value of $c$ is so small that
detecting the existence of the backdoor by simply guessing the value of the
designer key does not appear to be significantly easier than guessing one of the
factors of the corresponding semi-primes. Every test trial involves
choosing a value for the parameter $K$, generating a designer key $T$ and a pair of
vulnerable semi-primes, then recovering the factors of the semi-primes by just
using the values of the semi-primes and the designer key. We basically executed
the tests by varying the parameter $K$ so as to determine a value yielding both
fast generations of vulnerable semi-primes and reasonably quick recovery of the
factors.

\begin{table}
    \begin{filecontents*}{tsb512.csv}
K,GEN,GSD,REC,RSD
10,26.6,31.8,2.7,2.7
50,7.8,4.7,6.4,6.6
100,6.9,2.7,23.5,24.0
150,5.0,1.5,151.9,237.0
200,5.6,2.6,270.8,612.9
250,3.9,1.3,811.9,1277.7
300,3.9,1.3,1309.1,2681.1
350,4.2,1.2,1598.7,3494.7
400,4.5,2.3,2946.3,5371.8
\end{filecontents*}
\pgfplotstabletypeset[
      multicolumn names, % allows to have multicolumn names
      col sep=comma, % the separator in our .csv file
      display columns/0/.style={
        column name={$K$}, % name of first column
        column type=r},
      display columns/1/.style={
        column name={avg.},
        column type=r},
      display columns/2/.style={
        column name={st.dev.},
        column type=r,string type},
      display columns/3/.style={
        column name={avg.},
        column type=r,string type},
      display columns/4/.style={
        column name={st.dev.},
        column type=r,string type},
      every head row/.style={
        before row={\toprule
        \multicolumn{3}{l}{$\alpha{=}512$\hskip3.5ex Generation}&\multicolumn{2}{c}{Recovering}\\
        },
        after row=\midrule
      },
      every last row/.style={after row=\bottomrule} % rule at bottom
    ]{tsb512.csv} % filename/path to file
    \par\medskip
    \begin{filecontents*}{tsb1024.csv}
K,GEN,GSD,REC,RSD
10,454.7,402.5,9.8,8.0
50,95.3,69.7,23.2,20.1
100,59.3,34.4,75.1,81.3
150,57.8,35.2,133.2,286.4
200,43.5,23.9,315.6,641.3
250,45.8,21.8,981.4,2810.1
300,38.2,16.3,1905.0,3161.7
350,38.1,26.1,2388.7,5378.7
400,35.9,12.7,5695.4,11449.7
\end{filecontents*}
\pgfplotstabletypeset[
      multicolumn names, % allows to have multicolumn names
      col sep=comma, % the separator in our .csv file
      display columns/0/.style={
        column name={$K$}, % name of first column
        column type=r},
      display columns/1/.style={
        column name={avg.},
        column type=r},
      display columns/2/.style={
        column name={st.dev.},
        column type=r,string type},
      display columns/3/.style={
        column name={avg.},
        column type=r,string type},
      display columns/4/.style={
        column name={st.dev.},
        column type=r,string type},
      every head row/.style={
        before row={\toprule
        \multicolumn{3}{l}{$\alpha{=}1024$\hskip3.5ex Generation}&\multicolumn{2}{c}{Recovering}\\
        },
        after row=\midrule
      },
      every last row/.style={after row=\bottomrule} % rule at bottom
    ]{tsb1024.csv} % filename/path to file
    \par\medskip
    \begin{filecontents*}{tsb2048.csv}
K,GEN,GSD,REC,RSD
10,6353.2,3759.4,31.1,9.3
50,1785.4,1429.4,43.1,12.5
100,1086.0,887.3,104.4,108.7
150,647.1,376.5,236.3,290.4
200,544.3,277.6,619.9,729.6
250,456.8,305.3,1976.0,3493.5
300,395.4,236.6,1910.5,4716.5
350,407.6,155.3,2537.8,5460.6
400,321.1,140.0,4541.4,6038.2
\end{filecontents*}
\pgfplotstabletypeset[
      multicolumn names, % allows to have multicolumn names
      col sep=comma, % the separator in our .csv file
      display columns/0/.style={
        column name={$K$},
        column type=r},
      display columns/1/.style={
        column name={avg.},
        column type=r,string type},
      display columns/2/.style={
        column name={st.dev.},
        column type=r,string type},
      display columns/3/.style={
        column name={avg.},
        column type=r,string type},
      display columns/4/.style={
        column name={st.dev.},
        column type=r,string type},
      every head row/.style={
        before row={\toprule
        \multicolumn{3}{l}{$\alpha{=}2048$\hskip3.5ex Generation}&\multicolumn{2}{c}{Recovering}\\
        },
        after row=\midrule
      },
      every last row/.style={after row=\bottomrule} % rule at bottom
    ]{tsb2048.csv} % filename/path to file

    \caption{\TSB: running times (in seconds, average and standard
    deviations on 20 trials) for $\alpha=512, 1024, 2048$.} 
    \label{t:tsb}
\end{table}

\begin{figure}
\includegraphics[width=\columnwidth]{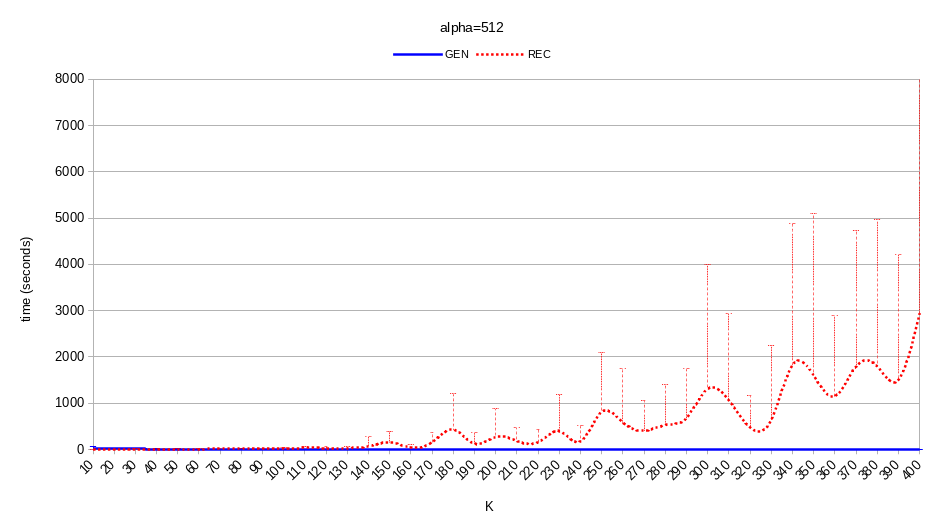}
\par\medskip
\includegraphics[width=\columnwidth]{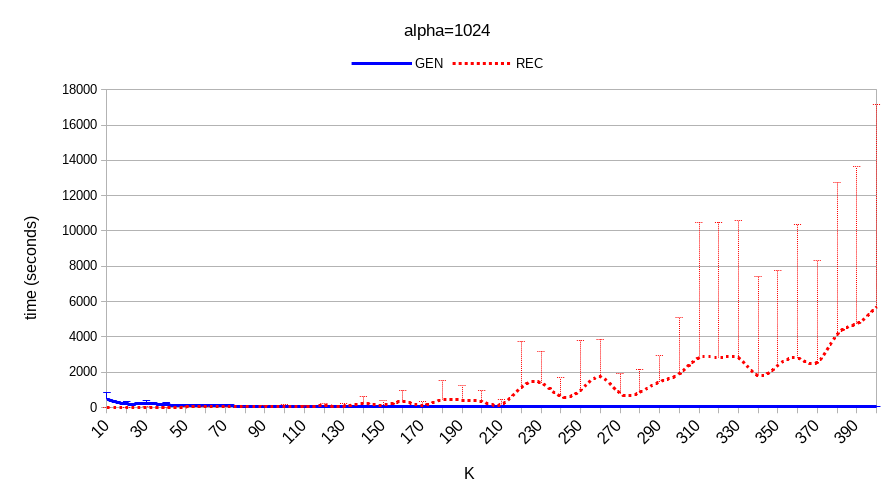}
\par\medskip
\includegraphics[width=\columnwidth]{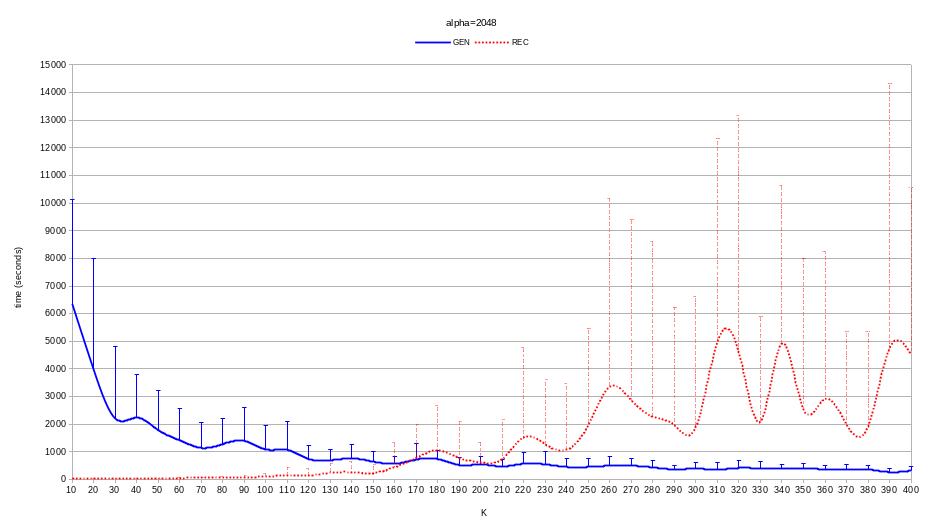}
\caption{\TSB: average running times for $\alpha=512, 1024, 2048$ (20 trials
for each value of $K$).  Magnitudes of the standard deviations are shown as
vertical bars.}
\label{f:tsb}
\end{figure}

The tests have been executed on the same computational nodes described in 
Section~\ref{s:ssb}.
All tests have properly recovered the factors of the vulnerable semi-primes.
Each value of $K\in\set{10\cdot i\,|\, i=1,\ldots, 40}$ has been tested 20
times. The SageMath code is sequential, that is, each test trial runs on a
single computation core. We report in Table~\ref{t:tsb} and Figure~\ref{f:tsb}
averages and standard deviations of the running times.

The value of $K$ is crucial in determining both the time required to generate a
pair of semi-primes and the time required to recover the factors. The
experimental results show that, even if the SageMath code is not optimzed, the
recovery time is reasonably small for all tested values of~$K$, hence \TSB\ is
a practically effective backdoor. However, generation time is also very
important whenever the backdoor mechanism has to be hidden in hardware devices
or software programs that are supposed to yield robust, legit semi-primes.
While in general larger values of~$K$ are associated to smaller generation
times, there seems to be a threshold value for $K$ above which the generation
times are essentially constants and near the minimum observed value. From the
values shown in Figure~\ref{s:tsb} we may safely set $K$ to values near $100$,
$200$, and $400$ for $\alpha=512$, $1024$, and $2048$, respectively, that is,
$K\approx\alpha/5$.

\section{Conclusions}
\label{s:conc}

We presented a new idea for designing backdoors in cryptographic systems based
on the integer factorization problem. The idea consists in introducing some
mathematical relations among the factors of the semi-primes based on
congruences modulo a large prime chosen by the designer. A first algorithm,
\SSB, can be used to implement a symmetric backdoor, hence the designer key
acts as a pure escrow key that must be kept hidden to the owner of the generated
keys (in order to hide the vulnerability) and to third-party attackers. Another
proposed algorithm, \TSB, injects a vulnerability in a pair of distinct
semi-primes and may be used to implement both a symmetric backdoor and an
asymmetric one. It is interesting to observe, however, that it does not seem to
be hard to plug an asymmetric cipher in both \SSB\ and \TSB, similarly to the
mechanism implemented in \cite{markelova2021} for the Anderson's backdoor; this
may be a future evolution of the present work.

We implemented both \SSB\ and \TSB\ in SageMath and conducted extensive
experiments to determine optimal values for a crucial parameter of the
algorithms, which basically sets a trade-off between the generation time
of the vulnerable semi-primes and the recovery time when exploiting the
backdoors. The SageMath code has not been optimized at all, however even
for large RSA-4096 keys the recovery time is reasonably small (a few hours, at
worst, on a single computation core).

A crucial point is minimizing the generation time of the vulnerable
semi-primes. Our generation algorithms are not very sophisticated or optimized,
because basically they generate random values in the hope to find the proper
primes satisfying the mathematical conditions of the backdoors. We would like
to get generation times similar to those of legit public key generators.
However, an analysis of the performances of semi-prime generators likely
depends on the characteristics of the underlining pseudo-random number
generators, which also may depend on external factors such as the amount of
entropy collected by the system (see for example Linux's PRNG). Such
analysis is not simple, hence it has to be deferred to a future work.

\ifdoubleblind\else
\subsection*{Acknowledgments}

We gratefully thank Epigenesys s.r.l. for providing the computational resources
used in the experimental evaluation of our algorithms. We also thank E.\@ Ingrassia
e P.\@ Santucci for their valuable comments and suggestions.
\fi

\subsection*{Declaration of interests}

The author declares that he have no known competing financial interests or
personal relationships that could have appeared to influence the work
reported in this paper.

This research did not receive any specific grant from funding agencies in the public,
commercial, or not-for-profit sectors.

\bibliography{refs}{}
\bibliographystyle{unsrt}
\end{document}